\input{psfig.sty}

\documentclass{aa}

\def\saxj{SAX J1808.4--3658}

\begin{document}

\title{The Optical Counterpart to \saxj\ in Quiescence: Evidence of an 
Active Radio Pulsar?}

  \subtitle{}

   \author{Luciano Burderi\inst{1} \and Tiziana Di Salvo\inst{2,3} 
           \and Francesca D'Antona\inst{1} \and Natale R. Robba\inst{3}
           \and Vincenzo Testa\inst{1}}
   
   \offprints{L. Burderi}

   \institute{Osservatorio Astronomico di Monteporzio, via Frascati 33, 
   00040 Roma, Italy.
              email: burderi@coma.mporzio.astro.it
         \and
              Astronomical Institute "Anton Pannekoek," University of 
              Amsterdam and Center for High-Energy Astrophysics,
              Kruislaan 403, NL 1098 SJ Amsterdam, the Netherlands.
              email: disalvo@astro.uva.nl
         \and
              Dipartimento di Scienze Fisiche ed Astronomiche, 
              Universit\`a di Palermo, via Archirafi 36 - 90123 Palermo, Italy}

   \date{Received ; accepted }
   
   \authorrunning{L. Burderi et al.}
   \titlerunning{On the Optical Counterpart to \saxj\ in Quiescence}

   \maketitle

\begin{abstract}
The optical counterpart of the binary millisecond X-ray pulsar
\saxj\ during quiescence was detected at V = 21.5 mag by 
Homer et al.\  (2001). This star shows a 6\% semi-amplitude
sinusoidal modulation of its flux at the orbital period of the system.
It was proposed that the modulation arises from X-ray 
irradiation of the intrinsically faint companion by  
a remnant accretion disk, and that the bulk of the optical emission 
arises from viscous dissipation in the disk.
The serious difficulty in this scenario lies in the estimate of the 
irradiating luminosity required to match the observational data, that 
is a factor $10-50$ higher than the quiescent X-ray
luminosity of this source.
To overcome this problem, we propose an alternative scenario, in
which the irradiation is due to the release of rotational energy by
the fast spinning neutron star, switched on, as magneto-dipole rotator
(radio pulsar), during quiescence. Our computations indicate that the 
optical magnitudes are fully consistent with this hypothesis. In this case 
the observed optical luminosity may be the first evidence that a radio 
pulsar is active in this system in quiescence, a key phase for understanding 
the evolution of this class of objects.
\end{abstract}


\section{Introduction}

\saxj\ is a transient low mass X-ray binary (hereafter LMXB) first detected 
with the Wide Field Cameras on board BeppoSAX in 1996 (in't Zand 
et al.\ 1998). The source showed type I X-ray bursts and was classified as a 
LMXB in which the compact object is a neutron star (hereafter NS). Analysis 
of the burst emission, showing Eddington-limited photospheric radius 
expansion, allowed an estimate of the source distance of 4 kpc (in't
Zand et al.\ 1998). This distance has been recently updated to 2.5 kpc
(in't Zand et al.\  2001). \saxj\ shows X-ray outbursts, lasting
a few tens of days, approximately every two years, with outburst luminosities 
between $10^{35}$ and $10^{36}$ ergs/s and quiescent luminosities between 
$10^{31}$ and $10^{32}$ ergs/s (see e.g.\ Wijnands et al.\ 2001; 
Campana et al.\ 2002).

In RXTE observations performed during the 1998 outburst 
Wijnands \& van der Klis (1998) detected coherent pulsations, at a frequency 
of $\sim 401$ Hz (corresponding to a NS spin period of $\sim 2.49$ ms).
Timing analysis on the same data
revealed Doppler orbital modulation with a period of $\sim 2.01$ hr and
negligible eccentricity. The X-ray flux shows a weak modulation of $\sim 2\%$ 
with a broad minimum when the NS is behind the companion. Adopting a typical 
NS mass of $M_1 = 1.35 M_{\odot}$, the mass function of the system suggests 
a very low mass companion, $M_2 \la 0.14 M_{\odot}$ at 95\% confidence level, 
of inconspicuous intrinsic luminosity (Chakrabarty \& Morgan 1998).  
\saxj\ was the first (low magnetized) 
LMXB to show coherent pulsations in the millisecond range in its persistent 
emission, thus providing the first evidence of the current evolutionary 
scenario, the so called recycling scenario, according to which LMXBs are 
the progenitors of binary millisecond radio pulsars. These are systems
containing a NS, active as a radio pulsar, orbiting a light 
($\sim 0.1 M_{\odot}$)
companion and spinning at millisecond periods, with a relatively weak
surface magnetic field ($\sim 10^8 - 10^9$ Gauss, as derived from the
measure of the spin period derivative). 

The optical counterpart of \saxj\ was identified by Roche et al.\ (1998)
with the variable star V4584 Sagittarii. Optical spectra of the companion
star showed absorption lines which are characteristic of mid to late type
stars (Filippenko et al.\ 1998), thus supporting the conclusion that the 
companion star has a very low mass (probably a low-mass, 
irradiation-bloated, brown dwarf; Bildsten \& Chakrabarty 2001).
Giles et al.\ (1999) reported on optical photometry
during the decline of the 1998 outburst (April-June): the source faded 
from $V = 16.72$ to 
$V \ga 20.50$.
The optical flux also showed a $0.06-0.08$ mag semi-amplitude modulation, 
roughly in antiphase with the weak X-ray orbital modulation, at a period 
which is consistent with the X-ray binary period of the system.
The phasing of the X-ray optical modulation is consistent with X-ray
reprocessing on the facing hemisphere of the companion star.
Wang et al.\ (2001) interpreted the observed optical flux in outburst as
emission from an X-ray heated accretion disk.

%%\subsection{The optical counterpart to \saxj\ in quiescence}

Homer et al.\ (2001) reported on high time resolution CCD photometry of this 
optical counterpart observed when the X-ray source was in quiescence. 
The optical component was detected at $V \sim 21.5$ mag, much fainter
than the value observed during the X-ray 1998 outburst. 
The two observations in quiescence reported by Homer et al.\ (2001)
were performed on 1999 August 10 and on 2000 July 3, respectively.  
An X-ray outburst of the source occurred 
between December 1999 and January 2000 and ended around May 2000 
(Wijnands et al.\ 2001). Therefore these observations
occurred roughly $4-5$ months before the start of this outburst 
and $1-2$ months after the end of the outburst, respectively. 
Colour images in B, V, and R obtained on 2000 July 3, were analyzed
and corrected for interstellar extinction towards \saxj, and are shown 
in Table~\ref{tab1}.
The ratio $F_X {\rm (2-10\; keV)} / F_{\rm V}$ is 
$\sim 1100$ in outburst and $\sim 10$ in quiescence, respectively.
This indicates that while the optical emission in outburst is certainly
due to X-ray reprocessing, similar to other outbursting transients,
X-ray reprocessing alone cannot be responsible of the bright optical  
emission observed in quiescence (Homer et al.\ 2001).

In these data a $\sim 6\%$ semi-amplitude modulation at the 2-hr orbital
period of the system is still significantly detected. The photometric minimum
is found when the companion star lies between the pulsar and the observer
and the shape of the modulation is approximately sinusoidal, similar to what 
is observed during outbursts. The faintness of the low-mass companion,
as well as the lack of double-humped morphology, due to an ellipsoidal 
modulation, excludes the direct optical emission from the companion as the
origin of the observed optical flux and modulation.

Homer et al.\ (2001) interpret the observed optical flux as due to viscous 
dissipation in an unirradiated accretion disk truncated at the corotation 
radius, $r_{\rm co} = (G M_{\rm NS} P_s^2/4 \pi^2)^{1/3}$ (where $G$ is the 
gravitational constant, $M_{\rm NS}$ is the NS mass and $P_s$ is the NS 
spin period), that is $\simeq 30$ km in the case of \saxj, and fuelled 
by a mass transfer rate of $\dot M \sim 10^{-11}\;M_\odot$/yr, consistent 
with gravitational radiation orbital angular momentum losses.
Although the adopted accretion rate is consistent with that measured during
X-ray outbursts (e.g.\ Wijnands et al.\ 2001), this disk model is only 
consistent with the lower limits on the optical fluxes, indicating that
the required mass accretion rate may be even higher than the one adopted.
The optical modulation is interpreted as due to X-ray irradiation of the 
companion star by the remnant accretion disk. From the amplitude modulation
the required level of irradiating flux has been estimated to be 
$L_{\rm irr} \sim 10^{33}$ ergs/s (Homer et al.\ 2001).

As already pointed out by Homer et al.\ (2001) this scenario meets with some 
difficulties. 
First of all, an X-ray luminosity is expected due to accretion onto the NS 
at a level of $\sim 10^{35}$ erg/s, adopting a standard efficiency of 
$\sim 10\%$ for accretion onto NSs. 
In principle a propeller effect could centrifugally inhibit 
accretion, although a residual $10^{33}$ erg/s are still needed to heat the 
companion at the level required from the optical data. This is a factor
of $10-50$ higher than the measured X-ray luminosity of \saxj\ in quiescence.
The quiescent X-ray luminosity of \saxj\ has been measured several
times, always resulting lower than $\sim 10^{32}$ ergs/s (Stella et al.\ 2000;
Dotani et al.\ 2000; Wijnands et al.\ 2002; Campana et al.\ 2002).
The most recent measure, $L_q \sim 5 \times 10^{31}$ erg/s in the 
0.5--10 keV range, was obtained using an XMM observation taken in March 2001 
(Campana et al.\ 2002). 

In this letter we show that these difficulties are solved if we make the case 
that actually the irradiation of the remnant disk and/or the companion star
is due to illumination from the {\it radio pulsar}, 
which may switch on during quiescence, once the magnetospheric radius is 
pushed beyond the light cylinder radius (see e.g.\ Burderi et al.\ 2001).
We suggest that the optical emission and modulation during quiescence is 
the first evidence that a magneto-dipole rotator (i.e.\ a radio pulsar) is 
active in this system during quiescence.

\section{Irradiation of the disk and companion star by rotating magneto-dipole 
emission}

It has been demonstrated that a rotating magnetic dipole in vacuum
emits electromagnetic dipole radiation. Also, a wind of relativistic 
particles associated with magnetospheric currents along the field lines
is expected to arise in a rotating NS (e.g.\ Goldreich \& Julian 1969). 
This radiative regime occurs when the space surrounding the NS
is free of matter up to the light cylinder radius ($r_{\rm LC} = c P_s/2 \pi$, 
where $c$ is the speed of light, i.e.\ the radius at which an object 
corotating with the NS attains the speed of light; see Ruderman et al.\ 
1989; Illarionov \& Sunyaev 1975).
Di Salvo \& Burderi (2003) have shown that a magneto-dipole rotator can 
easily be active in \saxj\ during quiescence. 
In fact, for a luminosity in quiescence of  
$\sim 5 \times 10^{31}$ ergs/s (Campana et al.\ 2002),
the NS surface magnetic field should be $B \la 0.05 \times 10^8$ 
Gauss in order to truncate the disk inside the corotation radius and
allow accretion onto the NS surface. This upper limit is incompatible with 
the presence of X-ray pulsations at the highest flux levels during outbursts,
which requires the disk to be truncated before the NS surface, and
therefore $B \ge 1 \times 10^{8}$ Gauss (Psaltis \& Chakrabarty 1999). 
On the other hand, the onset of a propeller regime or the intrinsic emission 
in X-rays by a magneto-dipole rotator switched-on during quiescence,
implies upper limits on the magnetic field above $1 \times 10^{8}$ Gauss, 
compatible with the presence of X-ray pulsations during outbursts.

In the hypothesis that the magneto-dipole rotator is active in \saxj\ during
quiescence, we can evaluate the power of the pulsar beam and, consequently 
the irradiation luminosity. To this aim we need an estimate of the NS 
magnetic field in this system.
As mentioned above, 
from a measure of the source luminosity in quiescence 
and using simple considerations on the position of the magnetospheric radius 
during quiescent periods, it is possible to estimate an upper limit on the 
NS magnetic field of $B \la 5 \times 10^8$ Gauss (Di Salvo \& Burderi 2003). 
This, together with the lower limit mentioned above, constrains the 
SAX J1808.4--3658 NS magnetic field in the quite narrow range 
$(1-5) \times 10^8$ Gauss.
Adopting this magnetic field we can calculate
the spin-down energy loss of the pulsar, i.e.\ the energy emitted by the
pulsar in the case in which the radio pulsar switches on:
$L_{PSR} = (2/3c^3) \mu^2 \omega^4 = 3.85 \times 10^{35} P_{-3}^{-4} 
\mu_{26}^2 \; {\rm ergs/s} \sim (1 - 25) \times 10^{34}$ ergs/s, 
where $\omega$ is the rotational frequency of the NS, $P_{-3}$
is the NS spin period in milliseconds, $\mu$ is the NS magnetic moment,
and $\mu_{26}$ is the NS magnetic moment in units of $10^{26}$ Gauss cm$^3$.

In the hypothesis that the pulsar beam heats a remnant accretion disk and/or 
a side of the companion star and assuming isotropic emission, we can evaluate 
the fraction of the irradiation luminosity that will be intercepted and 
reprocessed by the disk and the companion star, respectively. 
For the accretion disk this fraction, $f_D$, is given by the projected
area of the disk as seen by the central source, $2 \pi R \times 2 H(R)$ 
(where $R$ is the disk outer radius and $H(R)$ is the disk semi-thickness 
at $R$) divided by the total area, $4 \pi R^2$.
Adopting a standard Shakura-Sunyaev disk model 
(Shakura \& Sunyaev 1973),
we find: $f_D \simeq 1.6 \times 10^{-2} \alpha^{-1/10} \dot M_{-10}^{3/20} 
m_1^{-3/8} R_{10}^{1/8}$, where $\alpha$ is the viscosity parameter, 
$\dot M_{-10}$ is the mass accretion rate in units of $10^{-10}\; M_\odot$/yr,
$m_1$ is the NS mass in solar masses,
and $R_{10}$ is the outer radius of the disk in units of $10^{10}$ cm assumed 
to be at $\sim 0.8$ of the Roche lobe radius, $R_{L1}$, of the primary.
For a reasonable value of the viscosity parameter, $\alpha = 0.1$,
and adopting $R_{10} \simeq 2.9$ and $\dot M_{-10} = 0.1 - 1$ (note that 
$f_D$ depends weakly on the mass accretion rate), we find
$f_D \sim (1.5-2.1) \times 10^{-2}$. 

Indeed the outer accretion disk can still be present in the system during the 
optical observations performed in quiescence by Homer et al.\ (2001).
%%%in July 2000. 
In fact, the observations in July 2000 were performed approximately 
$1-2$ months from the end of the nearest outburst (May 2000). 
This time interval can be compared with the time scale to ``evaporate'' 
the disk, once the magneto-dipole rotator is active.
This time can be computed from the relation
$L_{PSR}\; f_D\; t_{\rm evap}= 0.5 <{U}_{D}> M_{D}$, where
$<{U}_{D}>$ and $M_{D}$ are the average potential and the mass of
the disk, respectively.
For \saxj\ this rough calculation gives $t_{\rm evap} \sim 30$ days.
On the other hand, the time scale to re-form the disk will be of the order 
of (or larger than) the viscous time scale of the disk: 
$t_{\rm visc} = 2.6 \times 10^5 \alpha^{-4/5} \dot M_{-10}^{-3/10} 
m_1^{1/4} R_{10}^{5/4}$ s. For \saxj\ $t_{\rm visc} \simeq 130$ days, that 
is comparable to the time interval between the optical observations performed
in August 1999 and the beginning of the following outburst (December 1999).
This means that a significant fraction of the disc could still be present
during the July 2000 and August 1999 observations.

A fraction $f_C$ of the pulsar spin-down luminosity will be reprocessed by 
the companion star and emitted in the optical band; in this case the 
intercepted fraction can be written as: 
$f_C = 2 \pi a^2 (1- \cos\theta)/(4 \pi a^2)$,
where $a$ is the the orbital separation and $\theta$ is the angle subtended
by the companion star as seen from the central source; if the companion
star fills its Roche lobe, this can be written as: $\sin \theta = R_{L2}/a$,
where $R_{L2}$ is the Roche lobe radius of the secondary and 
$R_{L2}/a = 0.49 q^{2/3}/[0.6 q^{2/3} + \ln (1 + q^{1/3})]$ (Eggleton 1983).
Assuming a mass ratio of $q = m_2/m_1 = 0.14/1.35$
(Chakrabarty \& Morgan 1998), where $m_2$ and $m_1$ are 
the masses of the companion and the NS in solar masses, respectively, we 
obtain: $f_C \sim 1.1 \times 10^{-2}$.

\begin{table}
\caption[]{Quiescent optical apparent (de-reddened) magnitudes in three 
different bands of V4580 Sagittarius. The measured values are from 
Homer et al.\ (2001).}
\label{tab1}
\footnotesize
\begin{center}
\begin{tabular}{c c c c c c}
\hline \hline
 Band        & \multicolumn{2}{c}{Measured} & \multicolumn{2}{c}{Predicted} \\
             &       Lower limit     & Upper limit & 
                   $\mu_{26} = 5$ & $\mu_{26} = 1$ \\\hline
                 $B_0$ & 17.7 & 21.0 & 17.9 & 21.8 \\
                 $V_0$ & 18.2 & 20.8 & 18.6 & 21.7 \\
                 $R_0$ & 17.7 & 21.0 & 18.7 & 21.4 \\
\hline
\end{tabular}
\end{center}
\end{table} 
If both the outer accretion disk and the companion star are reprocessing the 
pulsar spin-down luminosity, the total fraction of this luminosity that 
will be intercepted and reprocessed is: 
$f = f_D + f_C \sim 3.1 \times 10^{-2}$ (where we adopted 
$f_D \sim 2.0 \times 10^{-2}$), giving an optical reprocessed
luminosity of $\sim 3 \times 10^{32} \mu_{26}^2$ ergs/s. 
At a distance of 2.5 kpc, and adopting an average inclination angle 
$<i> = 50^\circ$ of the normal to the plane of the disk with respect 
to the line of sight, 
this corresponds to fluxes of $F_C \sim 1.4 \times 10^{-13} \mu_{26}^2$ 
ergs cm$^{-2}$ s$^{-1}$ and $F_D \sim 3.5 \times 10^{-13} \mu_{26}^2$ 
ergs cm$^{-2}$ s$^{-1}$ 
from the companion and the disk, respectively. 
The corresponding blackbody temperatures are estimated to
be $T_C \sim 5430 \; \mu_{26}^{1/2}$ K for the companion star and 
$T_D \sim 5080 \; \mu_{26}^{1/2}$ K for the irradiated surface of the disk.
From the sum of two blackbodies of temperatures $T_C$ and $T_D$ and
fluxes $F_C$ and $F_D$, respectively, we have calculated the predicted 
apparent magnitudes in three optical bands, which are perfectly in agreement 
with the measures reported by Homer et al.\ (2001) in the same optical
bands. The results are shown in Table~\ref{tab1}.

In our model, the optical modulation should again be caused by the heated 
side of the companion star, in agreement with the lack of ellipsoidal 
variations that should be expected in the intrinsic optical emission if 
the companion star fills its Roche lobe, as required in this system.

\section{Conclusions}

The optical counterpart of the binary millisecond X-ray pulsar
\saxj\ during quiescence was detected at $V \sim 21.5$ mag by 
Homer et al.\  (2001). This star shows a $\sim 6\%$ semi-amplitude
sinusoidal modulation of its flux at $\sim 2.01$ hr, i.e.\ at the
orbital period of \saxj.
The absence of a double peaked profile, expected in case of
ellipsoidal modulation, implies that the bulk of the optical flux does
not arise from the isothermal photosphere of the companion, that is
believed to fill its Roche lobe because of the presence of accretion 
episodes indicated by frequent X-ray outbursts, and suggest irradiation of the 
companion star. 
In order to explain these puzzling observations
we have proposed here that 
this irradiation is due to the release of 
rotational energy by the fast spinning NS, switched on as a
magneto-dipole rotator (radio pulsar) once the magnetospheric radius is 
pushed beyond the light cylinder radius during quiescence. Our computations
indicate that the optical magnitudes are fully consistent with this
hypothesis.  
The scenario we propose does not require a high mass accretion rate through
the disk nor a quiescent X-ray luminosity of $10^{33}$ ergs/s, because 
the luminosity for the reprocessing is supplied by the rotational energy of 
the pulsar, $L_{\rm PSR} \sim (1 - 25) \times 
10^{34}$ ergs/s.
Also, we can explain why a similar semi-amplitude of $\sim 6\%$ is observed
in the optical modulation both in outburst and in quiescence, since
the reprocessor geometry in our scenario does not substantially change
between outburst and quiescence.

The presence of an active magneto-dipole emitter during quiescence
in \saxj\, as required in this scenario, suggest that \saxj\ should 
show radio pulsations in quiescence.   
Indeed, despite thoroughly searched in radio during its X-ray quiescent phase, 
no pulsed radio emission has been detected from this source up to now. 
In particular Gaensler et al.\ (1999) found an upper limit on the 
radio emission from this source of 0.12 mJy at 1.4 GHz using the Australia 
Telescope Compact Array.
Burgay at al.\ (2003) found an upper limit of 0.98 mJy 
at 1.4 GHz searching for pulsed emission at the Parkes radiotelescope.
The lack of radio detection can be caused by the 
presence of a strong wind of matter emanating from the system, i.e.\ the mass 
released by the companion star swept away by the radiation pressure of the 
pulsar, as predicted in the so-called radio-ejection model (Burderi et al.\
2001; see also Burderi et al.\ 2002).  This means that \saxj\ may 
show radio pulsations in quiescence when observed at frequencies higher than 
the standard 1.4 GHz (the frequency at which radio pulsars are normally 
searched), where the free-free absorption is less severe, as suggested
by Ergma \& Antipova (1999).
We therefore suggest that the observed optical flux and modulation in 
quiescence may be the first evidence that a radio pulsar is active in this 
system when it is in quiescence.

\begin{acknowledgements}
We thank the anonymous referee for useful suggestions.
This work was partially supported by a grant from the
Italian Ministry of University and Research (MURST Cofin 2001)
and the Netherlands Organization for Scientific Research (NWO).
\end{acknowledgements}

\end{document}